\documentclass[12pt]{article}
\usepackage{amssymb}
\newcommand{\be}{\begin{equation}}
\newcommand{\ee}{\end{equation}}
\newcommand{\bea}{\begin{eqnarray}}
\newcommand{\eea}{\end{eqnarray}}
\newcommand{\lb}{\label}

\begin{document}
\begin{titlepage}

\begin{center}
{\large \bf Peculiar Relics from Primordial 
Black Holes in the Inflationary Paradigm}
\vskip 1cm
{\bf A.Barrau\dag\, D.Blais\ddag\, G.Boudoul\dag\, D.Polarski\ddag\ }
\vskip 0.4cm
\dag\ Laboratoire de Physique Subatomique et de Cosmologie,
 CNRS-IN2P3/UJF\\
 53, av des Martyrs, 38026 Grenoble Cedex, France 
\vskip 0.4cm
\ddag\ Laboratoire de Physique Math\'ematique et Th\'eorique,
 UMR 5825 CNRS\\
 Universit\'e de Montpellier II, 34095 Montpellier, France

\end{center}
\date{\today}
\vskip 2cm
\begin{center}
{\bf Abstract}
\end{center}
\begin{quote}
Depending on various assumptions on the energy scale of inflation and 
assuming a primordial power spectrum of a step-like structure, we 
explore new possibilities for Primordial Black Holes (PBH)
and Planck relics to contribute substantially to Cold Dark Matter in the 
Universe. A recently proposed possibility to produce Planck relics in
four-dimensional string gravity is considered in this framework. 
Possible experimental detection of PBHs through gravitational waves is   
also explored. We stress that inflation with a low energy scale, and also 
possibly when Planck relics are produced, leads unavoidably to relics 
originating from PBHs that are not effectively classical during their 
formation, rendering the usual formalism inadequate for them.
\end{quote}


\end{titlepage}

\section{Introduction}
The formation of Primordial Black Holes in the early stages of the universe 
is a generic feature and it is therefore interesting to study its 
cosmological consequences \cite{CH74}. 
Whatever the formation model, the Primordial Black Hole (PBH) spectrum must 
be in agreement with two types of constraints. The
first one is associated with evaporation: the density must be low enough 
so that physical effects due to the Hawking radiation do not
contradict any observed phenomena. They are based on the entropy per baryon,
the $n\bar{n}$ production at nucleosynthesis, 
the deuterium destruction, the
Helium-4 spallation \cite{liddle} \cite{carr94} and, finally, on the  nowadays 
observed gamma-ray \cite{carr98} and antiproton \cite{barrau02} spectra. 
Those constraints apply for initial PBH mass between $10^9~{\rm g}$
and $10^{15}~{\rm g}$, as these are the initial masses that can influence the 
above observations through their (Hawking) evaporation. Using the quantity 
$\beta$, which gives the probability that a region has the required density 
contrast to form a PBH at the horizon crossing time corresponding to the 
considered scale, the cosmic-ray constraints are the more stringent ones, 
leading approximately to $\beta (M_{PBH}=5\times 10^{14}~{\rm g})<10^{-26}$ 
\cite{barrau02}.
The second type of constraints is associated with the normalization of the 
spectrum  on cosmological scales probed by Cosmic Microwave Background (CMB) 
data. Whatever the considered power spectrum to form PBHs, it must generate 
a correct density contrast on COBE scales, i.e. large angular scales 
corresponding to the present Hubble radius scale. 

A pure scale-invariant Harrison-Zel'dovich power spectrum from the CMB scales 
up to very small scales would lead to a negligible amount of PBHs \cite{CL93},
\cite{carr94}. 
The only way to produce PBHs as a significant dark matter candidate 
is to increase the power on small scales without 
contradicting the observational data. A first attempt in this 
direction would be to allow for a tilt: $P(k)\propto k^n$ with $n>1$. 
Even without considering possible
inconsistencies with cosmic-ray data, the required value, around $n\approx1.3$
\cite{polarski1}, seems extremely disfavoured by the analysis of the most
recent CMB experiments: between $n\approx 0.91\pm 0.06$ \cite{map} (WMAP 
measurements: CMB + running spectral index) and $n\approx 1.04 \pm 0.12$ 
\cite{benoit} (Archeops measurements: CMB + $H_0$). A natural alternative 
is to boost power on small scales by means of a bump in the 
fluctuations power spectrum as suggested, for example, in \cite{polarski1}, 
\cite{JY98}. 
We follow here this idea and show that a (very) wide new parameter space can
be opened for dark matter when the energy scale of inflation is low enough. 
Furthermore, in contrast with previous models, in such a scenario
a simple step-like structure in the spectrum is enough to generate a large
quantity of PBHs without being in conflict with observations. Indeed, as 
a sufficient low energy scale allows the model to evade the $\gamma$-ray 
background constraints, no bump is required in the mass variance. 
Some observational probes through the emission 
of gravitational waves by coalescing PBHs are suggested to test this hypothesis.
The paper deals also with new models for Planck relics formation, based on 
a four-dimensional effective action in the framework of string gravity, to show 
that even for a high-energy scale inflation, PBH-induced dark matter is a viable 
candidate.
We stress that in some of these scenarios the production of (non-evaporated) 
PBHs from quantum fluctuations which are not highly squeezed and therefore not 
effectively classical \cite{PS96}, is unavoidable and cannot be handled with 
the usual formalism \cite{P01}. This will be the case when the energy scale of 
inflation is low enough so that the Hubble mass at the end of inflation, 
$M_{H,e}$, is larger than $M_*\approx 5\times 10^{14}$g, the mass of PBHs 
ending their evaporation at the present time, and also in high-scale inflation 
for Planck relics whose initial mass at formation is close to $M_{H,e}\ll M_*$.

\section{Inflation with a low energy scale}

\subsection{A pure step}

An important consequence of low scale inflation is the decrease of the 
reheating temperature. 
Though in practice the reheating scale can be much lower than the energy scale 
at the end of inflation, it will be enough for our purposes to make the simplifying 
assumption that the reheating is instantaneous. 
A low reheating temperature is required in order to avoid the possible overproduction 
of gravitinos \cite{khlopov}: $T_{RH}<10^8$ GeV. This value was even decreased to 
$4\times10^6$ GeV in some works based on Lithium abundance \cite{khlopovli}. 
This makes the horizon size at the end of inflation very 
large with an associated Hubble mass $M_{H} > 10^{16}$ g. This point is 
extremely important for PBH dark matter as it allows to avoid the main problem 
explained in e.g. \cite{polarski2}, namely the gamma-ray constraint which comes 
from the contribution to the $\gamma$-ray background of evaporated PBHs.
Then only the gravitational constraint, namely constraints on the present 
abundance of PBHs, would apply for PBH masses $M_{PBH}$ 
greater than $M_{H,e}$, the Hubble mass at the end of inflation. 

Previous work on the subject argued that one way to produce a significant amount 
of dark matter in the form of PBHs is to increase the mass variance in a well 
localized region so as to remain in agreement with the gamma-ray constraint. 
For example, using the Broken Scale Invariance 
(BSI) Starobinsky spectrum \cite{starobinsky}, it was shown that the 
oscillation in the power spectrum due to the jump in the derivative of the 
inflaton potential should produce a bump in the mass variance \cite{polarski2}. 
This slight increase in variance $\sigma_H^2$ can boost the PBH formation 
probability $\beta$ by more than ten orders of magnitude.
The resulting bump in the probability $\beta$ to form PBHs can yield
$\Omega_{PBH,0}\simeq 0.3$ for $5\times 10^{15}~{\rm g} < M_{PBH} < 10^{21}~{\rm g}$ 
with values of $p\approx 8\times 10^{-4}$ \cite{polarski2} where $p^2$ is the ratio 
of the power on large scales with respect to that on small scales.

Clearly, if the horizon mass $M_{H,e}$ at the end of inflation is larger than 
$M_*\approx 5\times 10^{14}~{\rm g}$, the initial
mass of a PBH whose lifetime if equal to the age of the Universe, the 
$\gamma$-ray and antiproton constraints as well as all the other constraints on 
smaller masses associated with evaporation are automatically evaded without any 
requirement about the shape of the fluctuation spectrum.
PBHs with masses above $M_*$ are nearly insensitive to the Hawking emission 
as the temperature $T=\hbar c^3/(8\pi GkM_{PBH})$ becomes smaller than the rest mass 
of any known massive field. An extremely wide mass range without constraint (except, 
to some extent, for microlensing upper limits) is therefore opened. 

The relative PBH abundance today is given by \cite{polarski1}
\be
\Omega_{PBH,0}(M_{PBH})\approx 1.3 \times 10^{17} \beta (M_{PBH})
           \left( \frac{10^{15}~{\rm g}}{M_{PBH}}    \right)^{\frac{1}{2}}~, 
\ee
for $h\approx 0.7$, and the subscript $0$ stands for the present-day value. 
The quantity $\beta$ is defined as
\begin{equation}
\label{beta}
  \beta(M_{H})=\frac{1}{\sqrt{2\pi}\,\sigma_H(t_{k})}~
           \int_{\delta_{min}}^{\delta_{max}}\,
           e^{-\frac{\delta^2}{2 \sigma_H^2(t_{k})}}\,\textrm{d}\delta
          \approx\frac{\sigma_H(t_{k})}{\sqrt{2\pi}\,\delta_{min}}
           e^{-\frac{\delta_{min}^2}{2 \sigma_H^2(t_{k})}}~,
\end{equation}
where $t_k$ is the horizon crossing time for the considered mode, $\delta$ is the
density contrast, $M_H$ is the Hubble mass at $t_k$ and $\sigma_H^2(t_k)\equiv
\sigma^2(R)|_{t_k}$.
For an accurate calculation, the mass variance 
$\sigma^2(R) \equiv <(\frac{\delta M}{M})^2_R>$ is computed with a top-hat window 
function $W_{TH}$ and $R=\frac{H^{-1}}{a}|_{t_k}$, using the expression \cite{polarski3} 
\be\lb{sigFW}
\sigma_H^2(t_{k})=\frac{8}{81\pi^2} \int_0^{\frac{k_e}{k}} x^3~F(kx)~
                           W^2_{TH}(c_sx)~W^2_{TH}(x)~\textrm{d}x~.
\ee
In (\ref{sigFW}), $k_e$ corresponds to the Hubble crossing scale at 
the end of inflation, $c_s^2=\frac{1}{3}$, the quantity $F(k)$ is defined through 
the equality $k^3 P(k,t_k)= \left (\frac{2}{3}\right )^4 F(k)$ for
$t_k < t_{eq}$ where $P(k,t)$ is the power spectrum of the primordial 
fluctuations \cite{P02},\cite{polarski3}.
However, in order to give a conservative estimate of the increase in power on small 
scales, we can assume following \cite{polarski1}, that a (possibly smoothed-out) 
jump occurs around some characteristic mass $M_s$, $M_{H,e}< M_s\ll M_H(t_{eq})$, 
{\it in the mass variance} spectrum, viz.
\be
\sigma_{H}^{COBE}= p~\sigma_{H}(t_k)\lb{COBE}~.
\ee
Normalizing the mass variance with the CMB large angular 
scales measurements, the increase in power and therefore the amplitude 
$p$ of the step can be estimated as a function of the horizon mass at the 
end of inflation $M_{H,e}$ as follows
\be
p\approx\frac{\sigma_{H}^{COBE}}{\delta_{min}}\sqrt{LW\left\{\frac{1.7\times
10^{34}}
    {2\pi \Omega_{PBH,0}^2}\left[ \frac{10^{15}~{\rm g}}{M_{H,e}} \right]\right\}}
\ee 
where $LW$ stands for the Lambert-W function (with $LW(xe^x)\equiv x)$. With 
$\Omega_{PBH,0}\approx\Omega_{m,0}\approx 0.3$, the numerical
estimates are : $p\approx 6.5\times 10^{-4}$ for $M_{H,e}=10^{15}~{\rm g}$,
$p\approx 5.5\times 10^{-4}$ for $M_{H,e}=10^{25}~{\rm g}$,
$p\approx 4.1\times 10^{-4}$ for $M_{H,e}=10^{35}~{\rm g}$. In these estimates 
we have taken $\delta_{min}=0.7$. 
%
%An exact calculation using (\cite{sigFW}) would give       
%$\sigma_H(t_k) = (10/9) \alpha \delta_H^{COBE}$ with $\alpha$ computed as in 
%\cite{polarski3} for a step in the primordial power spectrum itself.   
%
In principle, the reheating temperature can be as low as the MeV scale (the nucleosynthesis 
temperature), leading to huge horizon masses around $10^{38}~{\rm g}$. This can be 
considered as the upper limit for the low-mass cutoff of PBH spectra.
It is interesting that this corresponds to the highest viable PBH masses 
if CDM is made of PBHs \cite{S03}.   
It opens a very wide parameter space ($M_{H,e}$,$M_s$) for PBH dark
matter. Furthermore, if the reheating temperature is smaller than 1 GeV
($M_{H,e}\gg 10^{16}~{\rm g}$), PBHs could be one of the viable CDM candidates left, 
as supersymmetric dark matter cannot contribute substantially to dark matter 
\cite{fornengo}.

\subsection{Gravitational waves}
Probing PBHs as a dark matter candidate is experimentally is very difficult. 
As long as their masses are greater than
$10^{15}~{\rm g}$, black holes do not radiate and become really black. A decisive 
way to detect
them, and to observationally confirm or exclude this model, could be to look for
gravitational waves from coalescing PBHs. The maximum 
distance $R_{max}$ between the Earth and the binary
system compatible with the sensitivity of a given detector for a fixed PBH mass 
$M_{PBH}$ is
given by \cite{thorne}: 
\begin{equation}
\left(\frac{R_{max}}{20~{\rm Mpc}}\right)\approx 3.6\cdot10^{-21}h^{-1}_{SBmin}
\left(\frac{M_{PBH}}{{\rm M}_{\odot}}\right)^{\frac{5}{6}}
                       \left(\frac{\nu}{100~{\rm Hz}}\right)^{-\frac{1}{6}}~,
\end{equation}
where $h_{SBmin}$ is the minimum characteristic amplitude of a wave lying within
the detectot sensitivity and $\nu$ is the considered
frequency. 
To estimate the number $n(M_{PBH},R_{max})$ of PBHs inside
this sphere, we used an isothermal profile inside the Milky-Way halo
(for $R<150~{\rm kpc}$) which is given by~:
\begin{equation}
\rho(r,\psi)=\rho_{\odot}\frac{R_C^2+R_{\odot}^2}{R_C^2+R_{\odot}^2-2rR_{\odot}
cos\psi+r^2}
\end{equation}
where $\rho_{\odot}\approx 5\times10^{-25}~{\rm g}{\rm cm}^{-3}$ is the local halo mass 
density, $R_C\approx 3$ kpc is the core 
radius, $R_{\odot}\approx 8$ kpc is the
galactocentric distance, $r$ is the distance of the binary system to the Earth, $\psi$ is 
the angle between the considered point and the galactic center seen from the Earth, and
finally $\rho(r,\psi)$ is the halo mass density at coordinates $r,\psi$. This leads to:
\begin{equation}
n(M_{PBH},R_{max})=\frac{\pi\rho_{\odot}}{M_{PBH}}\frac{R_C^2+R_{\odot}^2}
{R_{\odot}}\int_0^{R_{max}} \ln\left\{
\frac{R_C^2+R_{\odot}^2+2rR_{\odot}+r^2}{R_C^2+R_{\odot}^2-2rR_{\odot}+r^2}\right\}rdr.
\end{equation}
For $R\gg 150~{\rm kpc}$, an average dark matter distribution with $\rho\approx
0.3\rho_c$ is assumed.
Finally the coalescence rate $f$ within this volume is computed under the natural
assumption that the distribution function of the PBHs comoving separation is uniform
and can be straitfowardly obtained from \cite{nakamura}:
\begin{equation}
f\approx 3\left( \frac{M_{PBH}}{{\rm M}_\odot} \right)^{\frac{5}{37}}\times
\frac{n(M_{PBH},R_{max})}{t_0}~,
\end{equation}
where $t_0$ is the age of the Universe. Gathering all those formulae and performing
numerical estimates shows that if PBHs have masses above $2\times 10^{-5} {\rm M}_{\odot}$, they
should generate more than one ``event'' per year in the VIRGO detector. If the LISA
frequencies and sensitivity are considered, the minimal mass decreases down to
$10^{-11} {\rm M}_{\odot}$. The interesting mass range probed covers then nearly fifteen 
orders of magnitude, though it also overlaps with microlensing data which 
exclude a significant contribution between $2\times 10^{-7}{\rm M}_{\odot}$ and 
$1 {\rm M}_{\odot}$ \cite{eros}. 

\subsection{``Quantum'' relics }  

A generic feature of low scale inflationary models is that all PBHs that are 
produced will survive and not evaporate. In particular, those PBHs that would 
form right after inflation correspond to fluctuations that were not (long 
enough) outside the Hubble radius and are therefore not highly squeezed.
Indeed, for the first scales which re-enter the Hubble radius, 
producing in these models PBHs with $M_{PBH}\gtrsim M_{H,e}$, the fluctuations are 
{\it not} highly squeezed. 
This is equivalent to saying that the decaying mode is still 
present and actually of the same order as the growing mode, and cannot be 
neglected.
Hence such low scale inflation models lead to the possible production of PBHs by 
inflationary fluctuations that cannot be considered as stochastic classical 
fluctuations, so the PBHs are not evaporated by today, in contrast to high scale 
inflation where $M_{H,e}\ll M_*$ and for which PBHs with mass 
$M_{PBH}\sim M_{H,e}$ have evaporated already during the (very) primordial 
stage of our Universe. 
Even if there is no significant increase in power on small scales, PBHs will 
be produced that cannot be described as classical objects {\it and} that will 
survive till the present time. We might therefore 
call them {\it quantum relics}. The intriguing point is not their 
abundance, which should be low, but rather the very nature of these 
surviving objects.

We want to illustrate the production of ``quantum relics'' with a concrete 
high energy physics inspired low scale inflationary model.
This production can take place, for example, in the phenemenological model considered 
in the first part of this Section.

It would be interesting if the scale of inflation corresponds to the supersymmetry 
breaking scale or even the electroweak scale. Initial 
conditions (through thermal effects) could set the inflaton field 
$\phi$ close to the origin where some symmetry is unbroken, as in 
``new inflation''. Inflation then takes place at small field values. 
At low temperature the inflaton starts rolling away from the 
origin $\phi=0$ (the effective mass term becomes negative),
spontaneously breaking the underlying symmetry. 
The following quite general inflationary potential 
can be considered in the context of supergravity inflation \cite{GRS99}:
\begin{equation}
V=\Lambda^4\times\left[\left(1-\kappa\frac{|\phi|^{p}}{\Lambda^qM^{p-q}_p}\right)^2+
\left(b+c\ln\left(\frac{|\phi|}{M_p}\right)\right)\left(\frac{|\phi|}{M_p}\right)^2\right]~,
\end{equation}
where $\Lambda$ is the near-constant vacuum energy driving inflation, $M_p$ the 
Planck mass, $b$ a ``bare'' mass term, $c$ a logarithmically varying mass term 
brought by radiative corrections, and $\kappa$, $p$ and $q$ determine the end of 
inflation by inclusion of higher-order terms.
%
%The scale of the inflationnary potential $\Lambda$ is very sensitive to the 
%form of the potential at the end of inflation
%and thus to the parameters $p$ and $q$. 
%
It has been shown that a value for $\Lambda$ from as low as $1$ GeV up to $\sim 10^{11}$ GeV
can be obtained for reasonable choices of $p$ and $q$, while still generating 
an acceptable spectrum of perturbations 
%
%$\delta_H\sim 2\times 10^{-5}$, 
%
with a spectral index $n_s\lesssim 0.95$ close to $1$ and a sufficient number of e-folds.
So the Hubble mass $M_{H,e}$ at the end of inflation will satisfy  $M_{H,e}>M_*$. 
We can quantify the degree of classicality of the formed PBHs 
with the ratio $D$ of the growing to the decaying mode for the scales under 
consideration \cite{P01}. A ratio $D\gg 1$ for a given scale corresponds to 
effective classicality of the fluctuations on this scale. 
We estimate it for adiabatic fluctuations in these models and we find 
towards the end of inflation, following \cite{P01}, the expression
\be
D(M) \simeq \left (\frac{\Lambda}{M}\right )^4~,\lb{D}
\ee
where $M^4$ stands for the energy density at the time when the PBH is formed. 
It is related to $M_H$ through 
\be
M_H = 5.6 \left(\frac{10^8~{\rm GeV}}{M} \right )^2 \times 10^{16} {\rm g}\lb{MH}~.
\ee  
The expression (\ref{D}) assumes that the inflationary energy density scale 
$\sim \Lambda^4$, and it applies to PBHs formed shortly after the inflationary phase.
As can be seen from (\ref{D}), only those PBHs produced immediately after the 
inflationary phase have $D\sim 1$ and are therefore quantum objects.
Though PBHs are here clearly irrelevant as CDM candidates because of their negligible 
abundance, the appearance of these ``quantum relics'' might be one of the few ways 
in which the quantum nature of the inflationary fluctuations can be exhibited. 
%%%%%%%%%%%%%%%%%%%%%%%%%%%%%%%%%%%%%%%%%%%%%%%%%%%%%%%%%%%%%%%%%%%%%%%%%%%%%%%%%%%%
%Finally, let us mention that a blue power spectrum, as an
%alternative to the step to increase the power on small scales, is not yet
%definitively ruled out. If reionisation was important enough to make the optical
%depth as high as 0.5, the measured spectral index of scalar fluctuations
%should be corrected in such a way that values up to more than 1.4 are not excluded 
%by a 99\% confidence level contour \cite{benoit}. Although quite extreme and
%unprobable this
%scenario is interesting as the required $n$ value to explain dark matter
%- around 1.32 \cite{polarski1} -
%for masses as low as $M_*$ will increase very slowly with $M_{RH}$. It is not
%increased by more than 3\% when $M_{RH}$ is increased from $10^{15}~{\rm g}$ 
%to $10^{35}~{\rm g}$. This possibility, which also escapes the cosmic-ray constraints
%in our model, becomes an appealing alternative to broken scale invariance.
%%%%%%%%%%%%%%%%%%%%%%%%%%%%%%%%%%%%%%%%%%%%%%%%%%%%%%%%%%%%%%%%%%%%%%%%%%%%%%%%%%%%
%%%%%%%%%%%%%%%%%%%%%%%%%%%%%%%%%%%%%%%%%%%%%%%%%%%%%%%%%%%%%%%%%%%%%%%%%%%%%%%%%%%%
%%%%%%%%%%%%%%%%%%%%%%%%%%%%%%%%%%%%%%%%%%%%%%%%%%%%%%%%%%%%%%%%%%%%%%%%%%%%%%%%%%%%
\section{Inflation with a high energy scale}

\subsection{Two distinct inflationary stages} 

Another way to evade the ``small scales'' problems for PBH formation still in 
the framework of high scale inflation is through 
the existence of a second inflationary stage at much lower energies.
So, a first stage of inflation solves all the problems usually solved by 
inflation, generates the cosmological perturbations observed, and produces 
also PBHs including in the ``dangerous'' mass interval around $M_*$.
However, due to the second stage of inflation, a significant 
PBH abundance produced during the first inflation is allowed.   
We give now in full generality the salient features of such a scenario.

Let us assume that the second inflation starts when the Hubble mass equals $M_{H,i}$, 
at a much lower energy than the first inflation scale. For simplicity we 
can assume the Hubble mass is constant during the second inflation.
%
%$M_{H,i}= M_{H,e}$. 
%
On a large range of scales, fluctuations produced during the first inflation will 
re-enter the Hubble radius, thereby possibly producing PBHs.
Part of those scales which re-entered the Hubble radius between the two inflationary 
stages will be expelled again 
outside the Hubble radius during the second inflation. Let us consider the scale 
$k_H\equiv a_iH_i$ that corresponds to the Hubble radius at the beginning of 
the second inflation. It will eventually, at the time $t_{k_H}$, reenter the 
Hubble radius when the Hubble mass is given by $M_H(t_{k_H})$. It is easy to 
derive the following relation between $M_H(t_{k_H})$ and $M_{H,i}$,
\be
\frac{M_H(t_{k_H})}{M_{H,i}} = e^{2N}~,    \lb{N}
\ee  
where $N$ is the number of e-folds during the second inflation. Due to a 
much lower energy scale, the amplitude of the produced fluctuations is 
quite negligible and will not lead to a significant PBH production. 
Only the fluctuations of the first, high scale, inflation will. 
Therefore there will be a gap in the mass 
range $M_{H,i}\leq M_{PBH} \leq M_H(t_{k_H})$. In addition, the density of all 
objects created before the second inflation will be reduced by an additional 
factor $e^{-3N}$. Hence, the only significant abundance of PBHs corresponds to 
the range 
\be
M_{PBH}\geq e^{2N} M_{H,i}~. \lb{delM}
\ee
Clearly it is possible to have PBHs as CDM in this range and still evade the 
small scales constraint coming from the evaporated PBHs.

An interesting low scale inflationary model of that kind is thermal inflation 
\cite{LS95}, triggered by a scalar field termed flaton, which can appear in 
SUSY theories. The consequences of thermal inflation on PBH abundance were
considered in \cite{GL97}.
By definition, a flaton has a large vacuum expectation value $M\gg 10^3$ GeV, 
while having a mass of order the electroweak  breaking scale 
$m\sim 10^2-10^3$ GeV. 
This leads to an almost flat potential for the flaton field 
$f$: $V\simeq V_0-m^2|f|^2$ with $V_0\simeq m^2M^2$. 
During thermal inflation the flaton field is held at the origin by finite 
temperature effects and the potential is dominated by the false vacuum energy 
$V_0$. Thermal inflation starts at the temperature $T_i\sim \sqrt{mM}$ 
when the thermal energy density falls below $V_0$ and ends at $T_e\sim m$ 
when the flaton can escape the false vacuum. The number of e-folds $N$ is then 
immediately given as $N=\frac{1}{2}\ln(\frac{M}{m})$ and the density of all 
PBHs produced before thermal inflation is suppressed accordingly by a factor 
$(\frac{m}{M})^{\frac{3}{2}}$. 
If we take $M\sim 10^{11}$ GeV, $m\sim 10^3$ GeV, thermal inflation starts at 
$T_i\sim 10^7$ GeV which corresponds to a Hubble mass 
$M_{H,i}\simeq 10^{18}~{\rm g}$, and ends at $T_e\sim 10^3$ GeV.
The number of e-folds is $N\approx 9$ and the density of PBHs with mass 
$M<M_{H,i}$ is suppressed by a factor $\sim 10^{-12}$. Clearly it is then possible 
to have a significant amount of PBHs and still evade the small scales constraints. 
Note however that with the numbers given above, PBHs as CDM can only exist in the range 
$M_{PBH}\gtrsim 10^{26}~{\rm g}$, starting near the edge of the range probed by the EROS 
data which constrain galactic dark matter in the range 
$2\times 10^{-7} {\rm M}_{\odot}\lesssim M_{PBH} \lesssim 1 {\rm M}_{\odot}$ 
\cite{eros}.
In particular, this could apply to the model with a step considered in the 
previous chapter, followed by a stage of thermal inflation.
However, a step in the primordial spectrum produced during the first period 
of inflation at the characteristic scale 
$M_s \lesssim 2\times 10^{-7} {\rm M}_{\odot}$ leaves possibly only a tiny 
mass interval ($M_H(t_{k_H})\lesssim M_{PBH}\lesssim 2\times 10^{-7} 
{\rm M}_{\odot}$) where PBHs are not constrained by observations. 
There will be essentially no ``quantum relics'' in this scenario as practically 
no PBHs are produced during the second (low-scale) inflation. 

\subsection{Planck relics} 

If we simply ignore moduli and gravitinos that appear in supersymmetric theories
and their possible overproduction if the reheating temperature is too high, 
as none of them as ever been detected, still another interesting possibility 
could be the production of Planck relics. Indeed, the unavoidable upper limit is imposed by
gravitational waves and fixes the smallest possible horizon mass after inflation
around $1~{\rm g}$. If the horizon mass is in this range, a natural way to produce dark matter
is through PBH relics. The idea was first mentioned in \cite{macgibbon}. Nevertheless,
two critical ingredients were missing at that time : the normalization of the primordial
spectrum to COBE data and a realistic (or, at least, possible) model to stop Hawking
evaporation in the Planck era. 
The latter point received a new light in the framework of
string gravity. The  following four-dimensional effective action with  second
order curvature corrections can be found:
\begin{equation}
S  =  \int d^4 x \sqrt{-g} \Biggl[ - R + 2 \partial_\mu \phi
      \partial^\mu \phi
   +  \lambda e^{-2\phi} S_{GB} + \ldots \Biggr],
\end{equation}
where $\lambda$ is the string coupling constant, $R$ is the Ricci scalar, $\phi$
is the dilatonic field and  $S_{GB} = R_{ijkl}R^{ijkl} -
4 R_{ij}R^{ij} + R^2$ is the curvature invariant -- the Gauss-Bonnet term. This
generalisation of the Einstein Lagrangian leads to the very important result 
that there is a minimal relic mass $M_{rel}$ for black holes produced in this 
model \cite{c07}.
Solving the equations at the first order of perturbation gives a minimal radius:
\begin{eqnarray}
r_h^{inf} = \sqrt{\lambda} \ \sqrt{4 \sqrt{6}}  \phi_h (\phi_\infty),
\end{eqnarray}
where $\phi_h (\phi_\infty)$ is the dilatonic value at the horizon. 
The crucial point is that this result remains true when higher 
order corrections or time perturbations are taken into account. The 
resulting value should be around $2 M_p$. It 
is even increased to $10 M_p$ if moduli fields are considered, making the 
conclusion very robust and conservative. 
The subsequent decrease of the Hawking evaporation leads to an
asymptotically stable state \cite{alexeyev}, giving a quantitative 
argument in favour of the existence of Planck relics. 
For formation masses above $10^9~{\rm g}$, important constraints 
are associated with Helium and Deuterium destruction. Actually, as 
we will see below, the relevant upper limit for the initial PBH mass 
is $\sim 10^5~{\rm g}$.

Once again, we consider a mass variance spectrum with a characteristic scale $M_s$. 
For small initial masses $M_{PBH}\sim 1{\rm ~g}$, the Planck relics relative 
density $\Omega_{rel,0}$ can be written as, cf.(1), 
\bea
\Omega_{rel,0} &\approx& 1.3 \times 10^{17} \beta (M_{PBH}) \frac{M_{rel}}{M_{PBH}} 
\left( \frac{10^{15}~{\rm g}}{M_{PBH}} \right)^{\frac{1}{2}}\\
&\approx& 2.83~\gamma \times 10^{-3} \beta(M _{PBH})
    \left( \frac{10^{15}~{\rm g}}{M_{PBH}} \right)^{\frac{3}{2}}~,\lb{Orel}
\eea
where we have taken $M_{rel}\equiv \gamma M_p$, while $M_{PBH}$ refers to the 
{\it initial} PBH mass. This leads to the following value for the step amplitude:
\begin{equation}
p\approx\frac{\sigma_{H}^{COBE}}{\delta_{min}}\sqrt{LW\left\{\frac{8.0\times
10^{-6}}{2\pi \Omega_{rel,0}^2}
\left[ \frac{M_{rel}}{M_p} \right]^2
\left[ \frac{10^{15}~{\rm g}}{M_{H,e}} \right]^3\right\}}.
\end{equation}
If Planck relics are to explain $\Omega_{m,0}\approx0.3$, $p$ varies 
from $7.1\times 10^{-4}$ to $5.5\times 10^{-4}$ for initial PBH masses between 
$1~{\rm g}$ and $10^5~{\rm g}$ with $M_{rel}=M_{p}$ and between 
$7.3\times 10^{-4}$ and $5.6 \times 10^{-4}$ with $M_{rel}=10M_{p}$. 
As for non-evaporating PBHs, $p$ has a very slight dependence on $M_{H,e}$ as 
$\beta$ is extremely sensitive to the mass variance in this range.

Equation (\ref{Orel}) is very accurate for small masses $M _{PBH}\simeq 1$~g 
and its validity extends up to 
$M _{PBH}\simeq M' \equiv 8~\gamma^{\frac{2}{5}}\times 10^5$~g,
with the corresponding range $10^{-21}~\gamma^{-1}\lesssim \beta(M _{PBH})
\lesssim 10^{-12}~\gamma^{-1}$.
We have checked this numerically using the simplifying assumption that all the
evaporation produces either a relativistic, or else a non-relativistic, component. 
For $M_{PBH}$ up to $M'$,
the energy density due to evaporation is still much smaller than the preexisting 
radiation background, at the time when the relics have formed. 
However, it should be stressed again that when $M_{PBH}\sim  M_{H,e}$, the quantity 
$\beta$ entering (\ref{Orel}) loses its meaning as a probability, not to mention 
the fact that the asymptotic domination of the growing mode is not achieved in this 
regime. 
Beyond $M'$, PBHs dominate the energy density before their evaporation is completed. 
This gives rise to a different expression for $\Omega_{rel,0}$ which is nearly independent 
of $\beta$ (\cite{carr94,barrow}).
Interestingly, in this mass range it is possible to obtain PBHs as CDM in significant 
amounts only for masses $M_{PBH}\simeq M'$. Thus, in the context of CDM, only 
equation (\ref{Orel}) is relevant and PBHs cannot contribute significantly to CDM if 
their mass lies in the range $M' \lesssim M_{PBH}\lesssim 10^9$~g.
Surprisingly, in such a scenario too, the notion of quantum relics resurfaces 
for those Planck relics originating from {\rm initial} PBH masses 
$M_{PBH}\simeq M_{H,e}\sim 1$~g. Because of their supposed large abundance, 
Planck relics with initial mass $M_{PBH}\sim M_{H,e}\sim 1$~g would probably 
be ruled out but we conjecture that a bump producing PBHs around $10^5$~g could 
yield a viable CDM candidate. 

\section{Conclusion}
In this work we have presented results of two different kinds. 
The first kind refers to the possibility of having PBHs as a viable CDM candidate.
We have shown that PBHs can be produced in significant amounts when the inflationary 
energy scale is low enough without being in conflict with constraints coming from 
the evaporation of PBHs with masses $M<M_*\approx 5\times 10^{14}$g, simply because 
in these models such PBHs do not form. On the other hand, if the inflationary scale is 
high, PBHs can still remain viable CDM candidates in theories where Planck relics exist, 
and we have considered this possibility too. Still another possibility is to have two 
inflationary stages, a high energy scale inflation followed by another 
inflation on a much lower energy scale. The possibility to detect PBHs through 
gravitational waves was also considered.

The second kind refers to the unavoidable production in low energy scale inflation
of PBHs that form from genuinely quantum-mechanical fluctuations and that are not evaporated 
by today. We have called these objects ``quantum relics''. 
As we have pointed out, such objects can also form in the context of Planck relics.
Clearly, the very existence of such objects poses a number of fascinating open problems 
both with respect to their formation and to their subsequent evolution and possible 
decoherence (see e.g. \cite{dec}).

\end{document}